\documentclass[12pt]{article}
\usepackage[dvips]{graphics}

\makeatletter
\def\dash{-}
\def\@citex[#1]#2{%
\if@filesw\immediate\write\@auxout{\string\citation{#2}}\fi
  \def\@citea{}\@cite{\@for\@citeb:=#2\do
     {\ifx\dash\@citeb{--}\def\@citea{}\else
      \@citea\def\@citea{,\penalty\@m\ }\@ifundefined
       {b@\@citeb}{{\bf ?}\@warning
       {Citation `\@citeb' on page \thepage \space undefined}}\fi
\hbox{\csname b@\@citeb\endcsname}}}{#1}}
\newbox\tempboxa
\newdimen\captionboxsubcount 

\newdimen\captionboxsub
\captionboxsub=\hsize \advance\captionboxsub by -\captionboxsubcount
\advance\captionboxsub by -\captionboxsubcount
\long\def\@makecaption#1#2{
 \setbox\@tempboxa\hbox{#1: #2}
 \ifdim \wd\@tempboxa >\captionboxsub 
\rightskip=\captionboxsubcount \leftskip=\captionboxsubcount #1: #2 
\else \hbox to\hsize{\hfil\box\@tempboxa\hfil} 
 \fi}
\makeatother
 
\renewcommand{\baselinestretch}{1.5}
 
\begin{document}
\setlength{\baselineskip}{16pt}


\thispagestyle{empty}

\rightline{OCHA-PP-138}
\rightline{DPNU-99-24}

\vspace{1cm}

\centerline
{\large\bf  On the $\eta^{\prime}$ Gluonic Admixture}

 \vskip2.0truecm
\renewcommand{\thefootnote}{\fnsymbol{footnote}}
\centerline{\large E. Kou
\footnote{ e-mail address: kou@eken.phys.nagoya-u.ac.jp}}
\bigskip
\centerline{\sl Dept. of Physics, Ochanomizu University, Tokyo 112-0012,
Japan}        
\centerline{\sl and}        
\centerline{\sl Physics Dept., Nagoya University, Nagoya 464-8602,
Japan}        
 \vskip0.5truecm 

\vspace*{3.5truecm}
\bigskip
\centerline{\large\bf Abstract } 

  The $\eta^{\prime}$ which is an $SU(3)_F$ singlet state  
  can contain a pure gluon component, gluonium. We examine this possibility 
  by analysing all available experimental data. 
  It is pointed out that the $\eta^{\prime}$ gluonic component 
  may be as large as 26\%. We also show that the amplitude for 
  $J/\psi\to\eta^{\prime}\gamma$ decay obtains a notable contribution 
  from gluonium.


%
%
%
%
%
%
%
%
%
%
%

\newpage
\renewcommand{\baselinestretch}{2.0}

\pagenumbering{arabic}


\section{Introduction}

The CLEO collaboration reported an unexpectedly large branching 
ratio for $B \to \eta^{\prime} X_s$ ~\cite{cleo}. 
One of the suggested mechanisms 
~\cite{bsgg1,-,other9}
to explain this problem 
considers the process $b \to sg$, $g \to
\eta^{\prime} g$
~\cite{bsgg1,-,bsgg9}. 
This mechanism is based on the anomalous coupling of 
$gg \to \eta^{\prime}$ which accounts for 
the large branching ratio for $J/\psi \to \eta^{\prime} \gamma$ decay. 
It should be noted that the gluonic component of $\eta^{\prime}$ 
has been studied extensively in the literature ~\cite{etap1,-,etap9}. 
We shall determine the gluonic component of $\eta^{\prime}$ 
considering all known experimental data. 

It is believed that $\eta^{\prime}$ consists of the $SU(3)_F$
singlet and octet
$q\bar{q}$ states which we denote as $\eta_1$ and $\eta_8$, respectively, and
dominated by the singlet state. The $SU(3)_F$ singlet state, differing from
the octet state, can be composed of pure gluon states. Therefore, we
examine another singlet state in $\eta^{\prime}$  
made only of gluons, which we call gluonium. 

The remainder of the paper is organized as follows. 
In Section 2, we describe our notation and introduce the gluonic
component. The formalism for studying the radiative light meson decays 
is presented in Section 3.
The recent discussions on the definition of the decay constants for 
$\eta$ and $\eta^{\prime}$
~\cite{Leut,-,Kroll9} are taken into account. 
We then proceed to obtain the pseudoscalar mixing angle $\theta_p$ and 
the possible gluonic content of $\eta^{\prime}$ in Section 4. 
The investigation of 
the radiative $J/\psi$ decay is performed in Section 5. 
A summary and conclusions are given in Section 6.


\renewcommand{\baselinestretch}{2.0}

\section{Notation} 

$SU(3)_F \times U(1)$ symmetry introduces the pseudoscalar octet state  
$\eta_8$ and singlet state $\eta_1$ as 
\begin{equation} 
        \left(\begin{array}{@{\,}c@{\,}}
        \eta_8 \\ \eta_1 \end{array} \right) 
   = \left(\begin{array}{@{\,}cccc@{\,}}
                                \sin{\theta_I} & -\cos{\theta_I} \\
                                \cos{\theta_I} & \sin{\theta_I} \\
                         \end{array} \right) 
           \left(\begin{array}{@{\,}c@{\,}}
           \frac{u\bar{u}+d\bar{d}}{\sqrt{2}} \\ 
            s\bar{s} \end{array} \right)
\label{def81}
\end{equation}
where $\theta_I$ is the ideal mixing angle which satisfies 
$\theta_I= \tan^{-1}{\frac{1}{\sqrt{2}}}$. 
The two physical states $\eta$ and $\eta^{\prime}$ are considered 
as mixtures of these states 
with pseudoscalar mixing angle $\theta_p$ 
\begin{equation} 
        \left(\begin{array}{@{\,}c@{\,}}
        \eta \\ \eta^{\prime} \end{array} \right) 
   = \left(\begin{array}{@{\,}cccc@{\,}}
                                \cos{\theta_p} & -\sin{\theta_p} \\
                                \sin{\theta_p} & \cos{\theta_p} \\
                         \end{array} \right) 
           \left(\begin{array}{@{\,}c@{\,}}
           \eta_8 \\ \eta_1 \end{array} \right) \label{mix}  
\label{deftp}
\end{equation}
Combining Eqs. (\ref{def81}) and (\ref{deftp}), we rewrite 
\begin{equation} 
        \left(\begin{array}{@{\,}c@{\,}}
        \eta \\ \eta^{\prime} \end{array} \right) 
   = \left(\begin{array}{@{\,}cccc@{\,}}
                                \cos{\alpha_p} & -\sin{\alpha_p} \\
                                \sin{\alpha_p} & \cos{\alpha_p} \\
                         \end{array} \right) 
           \left(\begin{array}{@{\,}c@{\,}}
           \frac{u\bar{u}+d\bar{d}}{\sqrt{2}} \\ 
            s\bar{s} \end{array} \right)  
\label{defalph}
\end{equation}
with $\alpha_p=\theta_p-\theta_I+\frac{\pi}{2}$ which 
represents the discrepancy of the mixing angle from the ideal one.
Note that the $\phi$ and $\omega$ in the vector meson system 
mix almost ideally, that is, $\alpha_v \simeq 0$ . 
This characteristic deviation from the ideal mixing 
in $\eta-\eta^{\prime}$ system can be understood in terms of the anomaly. 
Let us take the derivative of the singlet axial vector current 
\begin{equation}
\partial_{\mu}j^{\mu5}=2imq  \gamma_{5} \bar{q}
-\frac{3\alpha_s}{4 \pi}G_{\alpha \beta}
\tilde{G}^{\alpha \beta} \label{ano}
\end{equation}
where $G_{\alpha \beta}$ is a gluonic field strength and 
$\tilde{G}^{\alpha \beta}$ is its dual. 
The term proportional to $G\tilde{G}$ is coming 
from the triangle anomaly ~\cite{Adler}. 
It affects neither the octet axial vector nor the vector current. 
Eq. (\ref{ano}) implies that the pseudoscalar singlet state 
can be composed not only of $q\bar{q}$ but also of gluons. 
Treating the gluon composite equivalent to the quark composite, 
the $\eta^{\prime}$ which is mostly $SU(3)_F$ singlet 
may contain the pure gluon state, gluonium. 
Therefore, we reconstruct $\eta-\eta^{\prime}$ system by including
gluonium. 
Then Eq. (\ref{deftp}) is extended to 
a $3\times3$ matrix with 3 mixing angles 
\[   {\footnotesize
        \left(\begin{array}{@{\,}c@{\,}}
        \eta \\ \eta^{\prime} \\ i \end{array} \right)
   = \left(\begin{array}{@{\,}cccc@{\,}}
           \cos{\theta_p}\cos{\gamma}+\sin{\theta_p}\cos{\phi}\sin{\gamma} 
           &-\sin{\theta_p}\cos{\gamma}+\cos{\theta_p}\cos{\phi}\sin{\gamma} 
           &\sin{\phi}\sin{\gamma}\\
           \cos{\theta_p}\sin{\gamma}+\sin{\theta_p}\cos{\phi} 
           &\sin{\theta_p}\sin{\gamma}+\cos{\theta_p}\cos{\phi}\cos{\gamma}
           &\sin{\phi}\cos{\gamma} \\
           -\sin{\theta_p}\sin{\phi} 
           &-\cos{\theta_p}\sin{\phi} 
           & \cos{\phi} \\
                                 \end{array} \right) 
           \left(\begin{array}{@{\,}c@{\,}}
           \eta_8 \\ \eta_1 \\ gluonium \end{array} \right) 
} \]
where $i$ is a ''glueball-like state" which we refrain from discussing here. 
Since the mass of $\eta$ is about the mass of $\eta_8$ 
which is obtained from Gell-Mann Okubo mass formula, 
we assume that $\eta$ does not contain the extra singlet state gluonium. 
Setting $\gamma=0$, we obtain
\begin{equation}
        \left(\begin{array}{@{\,}c@{\,}}
        \eta \\ \eta^{\prime} \\ i \end{array} \right)
   = \left(\begin{array}{@{\,}cccc@{\,}}
                                \cos{\theta_p} & -\sin{\theta_p} & 0\\
                                \sin{\theta_p}\cos{\phi} & \cos{\theta_p}\cos{\phi} & \sin{\phi} \\
                          -\sin{\theta_p}\sin{\phi} & -\cos{\theta_p}\sin{\phi} & \cos{\phi} \\
                                 \end{array} \right) 
           \left(\begin{array}{@{\,}c@{\,}}
           \eta_8 \\ \eta_1 \\ gluonium \end{array} \right).
\end{equation}

It is convenient to write the $\eta$ and $\eta^{\prime}$ states 
as ~\cite{etap1} 
\begin{eqnarray}
        |\eta>&=&X_{\eta}|\frac{u\bar{u}+d\bar{d}}{\sqrt{2}}>+
        Y_{\eta}|s\bar{s}> \\ 
  |\eta^{\prime}>&=&X_{\eta^{\prime}}|\frac{u\bar{u}+d\bar{d}}{\sqrt{2}}>+
  Y_{\eta^{\prime}}|s\bar{s}> + Z_{\eta^{\prime}}|gluonium>. 
\end{eqnarray}
$X_{\eta (\eta^{\prime})}$, $Y_{\eta (\eta^{\prime})}$ and 
$Z_{\eta^{\prime}}$ are normalized as 
\begin{eqnarray}
  X_{\eta}^2 + \ Y_{\eta}^2=1  \label{X+Y=1}\\ 
  X_{\eta^{\prime}}^2 + \ Y_{\eta^{\prime}}^2+Z_{\eta^{\prime}}^2=1  
\end{eqnarray}
and relate to the mixing angles 
\begin{eqnarray}
   X_{\eta} &=&\cos{\alpha_p}, 
   \ \ \ \ \ \ \ \ \ \  Y_{\eta}=-\sin{\alpha_p},  \label{defy} \\
   X_{\eta^{\prime}}&=&\cos{\phi}\sin{\alpha_p}, \ \ \ 
   Y_{\eta^{\prime}}=\cos{\phi}\cos{\alpha_p}, \ \ \ \ \ 
   Z_{\eta^{\prime}}=\sin{\phi}. \label{defz} 
\end{eqnarray}                                                         


\renewcommand{\baselinestretch}{2.0}

\section{Decay rates} 

\vspace{0.5cm}

We calculate the decay rates by using the 
vector meson dominance model (VDM)
and the $SU(3)_F$ quark model (see for example, ~\cite{bramon,-,odonnell}).
In this method, the decay rates are expressed in terms of the masses and
the decay constants of light mesons.
The decay constants for vector mesons which are defined by 
\begin{equation}
m_Vf_V\epsilon^{\mu} = \langle 0|j^{\mu}_V|V(p,\lambda) \rangle \label{vdc}
\end{equation}
are well determined by their decays into $e^+e^-$ ~\cite{pdg} as 
\begin{equation}
 f_{\rho}=(216\pm5)\mbox{MeV}, \ \ \ f_{\omega}=(195\pm3)\mbox{MeV}, \ \ \
 f_{\phi}=(237\pm4)\mbox{MeV}.
\end{equation}
On the other hand, the decay constants for $\eta$ and $\eta^{\prime}$ 
are not well-defined because of the anomaly. 
Recently, there has been considerable progress 
on the parametrization of the decay constants of $\eta-\eta^{\prime}$ system 
~\cite{Leut,-,Kroll9}. 
Following Reference ~\cite{Kroll9}, 
we utilize the decay constants defined by 
\begin{eqnarray}
if_xp_{\mu} &=& \langle 0|u\gamma^{\mu}\gamma_5\bar{u}+d\gamma^{\mu}\gamma_5\bar{d}|
\frac{u\bar{u}+d\bar{d}}{\sqrt{2}} \rangle \label{deffx} \\
if_yp_{\mu} &=& \langle0|s\gamma^{\mu}\gamma_5\bar{s}|s\bar{s} \rangle \label{deffy} 
\end{eqnarray}
which are considered as the decay constants
for the $SU(3)_F$ singlet states at non-anomaly limit.
Since the state $|\frac{u\bar{u}+d\bar{d}}{\sqrt{2}} \rangle$
in Eq. (\ref{deffx}) is equivalent to $\pi^{0}$ but
an isospin singlet, we can approximately have the following relation
by assuming that the isospin breaking effect is not large: 
\[f_x=f_{\pi}.\]
When $SU(3)_F$ symmetry is exact
$f_y$ in Eq. (\ref{deffy}) is equal to $f_x$. 
However, the mass difference between
the $u$ and $d$ quarks and the $s$ quark is notable.
The Gell-Mann-Okubo mass formula gives a quantitative estimate of the 
$s$ quark mass breaking effect. 
Similarly, this breaking effect for our decay constants 
can be included through 
\[f_y=\sqrt{2f_K^2-f_{\pi}^2}.\]
The known values for $f_{\pi}=131~\mbox{MeV}$ and $f_{K}=160~\mbox{MeV}$ lead to
\begin{equation}
 f_x=131~\mbox{MeV}, \ \ \ f_y=1.41\times131~\mbox{MeV}.
\label{fxfyvalue}
\end{equation}
It is shown in Reference ~\cite{Kroll9} that 
the approximate values in Eq. (\ref{fxfyvalue}) are justified 
phenomenologically and also satisfy the result of 
chiral perturbation theory in ~\cite{Leut}. 

Using these decay constants, the radiative decay rates of 
the light mesons can be written in terms of $X_{\eta (\eta^{\prime})}$, 
$Y_{\eta (\eta^{\prime})}$ and $Z_{\eta^{\prime}}$ in the VDM as follows, 
\begin{eqnarray}
\Gamma(\omega\to\eta\gamma) &=& 
 \frac{\alpha}{24}\left(\frac{m_{\omega}^2-m_{\eta}^2}{m_{\omega}}\right)^3
 \left(\frac{m_{\omega}}{f_{\omega}\pi^2}\right)^2
\left(\frac{X_{\eta}}{4f_x}\right)^2  
\ \ \ \ \ \ \ \ \label{form1} \\
\Gamma(\phi\to\eta\gamma)   &=& 
 \frac{\alpha}{24}\left(\frac{m_{\phi}^2-m_{\eta}^2}{m_{\phi}}\right)^3
 \left(\frac{m_{\phi}}{f_{\phi}\pi^2}\right)^2
 \left(-2\frac{Y_{\eta}}{4f_y}\right)^2  
\ \ \ \ \ \ \ \  \label{form2} \\
\Gamma(\eta\to\gamma\gamma) &=& 
 \frac{\alpha^2}{288\pi^3}m_{\eta}^3
 \left(\frac{5X_{\eta}}{f_x}+\frac{\sqrt{2}Y_{\eta}}{f_y}\right)^2 
\ \ \ \ \ \ \ \  \label{form3} \\
\Gamma(\eta^{\prime}\to\omega\gamma) &=& 
 \frac{\alpha}{8}\left(\frac{m_{\eta^{\prime}}^2-m_{\omega}^2}
 {m_{\eta^{\prime}}}\right)^3
 \left(\frac{m_{\omega}}{f_{\omega}\pi^2}\right)^2
 \left(\frac{X_{\eta^{\prime}}}{4f_x}\right)^2 
\ \ \ \ \ \ \ \  \label{form4} \\
\Gamma(\eta^{\prime}\to\rho\gamma)   &=& 
 \frac{\alpha}{8}\left(\frac{m_{\eta^{\prime}}^2-m_{\rho}^2}
 {m_{\eta^{\prime}}}\right)^3
 \left(\frac{m_{\rho}}{f_{\rho}\pi^2}\right)^2
 \left(\frac{3X_{\eta^{\prime}}}{4f_x}\right)^2 
\ \ \ \ \ \ \ \  \label{form5} \\ 
\Gamma(\phi\to\eta^{\prime}\gamma)   &=& 
 \frac{\alpha}{24}\left(\frac{m_{\phi}^2-m_{\eta^{\prime}}^2}
 {m_{\phi}}\right)^3
 \left(\frac{m_{\phi}}{f_{\phi}\pi^2}\right)^2
 \left(-2\frac{Y_{\eta^{\prime}}}{4f_y}\right)^2  
\ \ \ \ \ \ \ \  \label{form6} \\ 
\Gamma(\eta^{\prime}\to\gamma\gamma) &=& 
 \frac{\alpha^2}{288\pi^3}m_{\eta^{\prime}}^3
 \left(\frac{5X_{\eta^{\prime}}}{f_x}
 +\frac{\sqrt{2}Y_{\eta^{\prime}}}{f_y}\right)^2  \label{form7}
\ \ \ \ \ \ \ \ 
\end{eqnarray}
where the OZI suppressed process occurring from 
$\phi-\omega$ mixing violation is ignored. 
In fact this breaking effect is expected to be very small; 
for example, in the case of the $\phi\to\pi^{0} \gamma$ decay, sin$\alpha_V$ 
is estimated to be less than 0.02. 

It is known that the VDM works quite well 
in the describing decay modes 
(see, for example, Refs. ~\cite{schechter,-,pancheri}). 
This is supported by performing the computation of the decay rates 
$\omega\to\pi^{0}\gamma $ and $\pi^{0}\to\gamma\gamma $ 
which do not depend on $X_{\eta (\eta^{\prime})}$, 
$Y_{\eta (\eta^{\prime})}$ and $Z_{\eta^{\prime}}$:  
\begin{eqnarray}
\Gamma(\omega\to\pi^{0}\gamma) &=& 
 \frac{\alpha}{24}\left(\frac{m_{\omega}^2-m_{\pi^{0}}^2}{m_{\omega}}\right)^3
 \left(\frac{m_{\omega}}{f_{\omega}\pi^2}\right)^2
\left(\frac{3}{4f_{\pi^{0}}}\right)^2 =0.72~\mbox{MeV}
\ \ \ \ \ \ \ \ \label{form8} \\
\Gamma(\pi^{0}\to\gamma\gamma) &=& 
 \frac{\alpha^2}{288\pi^3}m_{\pi^{0}}^3
 \left(\frac{3}{f_{\pi^{0}}}\right)^2 =0.0077~\mbox{KeV}. 
\ \ \ \ \ \ \ \  \label{form9} \ \ 
\end{eqnarray}
which are rather consistent with the experimental data ~\cite{pdg}
\[\ \ \ \ \ \ 
\Gamma(\omega\to\pi^{0}\gamma)=(0.72\pm0.043)~\mbox{MeV},
\ \ \ \ \ \ \ \\ \ \ \]
\[\Gamma(\pi^{0}\to\gamma\gamma)=(0.0077\pm0.00055)~\mbox{KeV}, \] 
respectively. Here we used $f_{\pi^{0}}=131~\mbox{MeV}$. 
In the case of the $\rho^{0}\to\pi^{0}\gamma$ decay, 
the model calculation gives 
$\Gamma(\rho^{0}\to\pi^{0}\gamma)=0.06~\mbox{MeV}$ 
which is small compared to the experimental value 
$\Gamma(\rho^{0}\to\pi^{0}\gamma)=(0.10\pm0.026)~\mbox{MeV}$. 
We note, however, that $\rho^{0}\to\pi^{0}\gamma$ decay rate  
still has a large error. 
It would be discussed in detail as more data will be available. 
We expect that the theoretical uncertainty occurring from 
the VDM is less than 15\%. 
This number is within the range of the error estimated in ~\cite{shifman} 
according to a QCD-based method.


\renewcommand{\baselinestretch}{2.0}
\vspace*{-0.5cm}
\section{Results}
\vspace*{-0.5cm}
\subsection{Results for $X_{\eta}$ and $Y_{\eta}$ 
(determination of $\theta_p$)}
\label{sec41}
\begin{figure}[tbh]
 \begin{center}
\vspace*{-4cm}
\includegraphics[\linewidth,\linewidth]{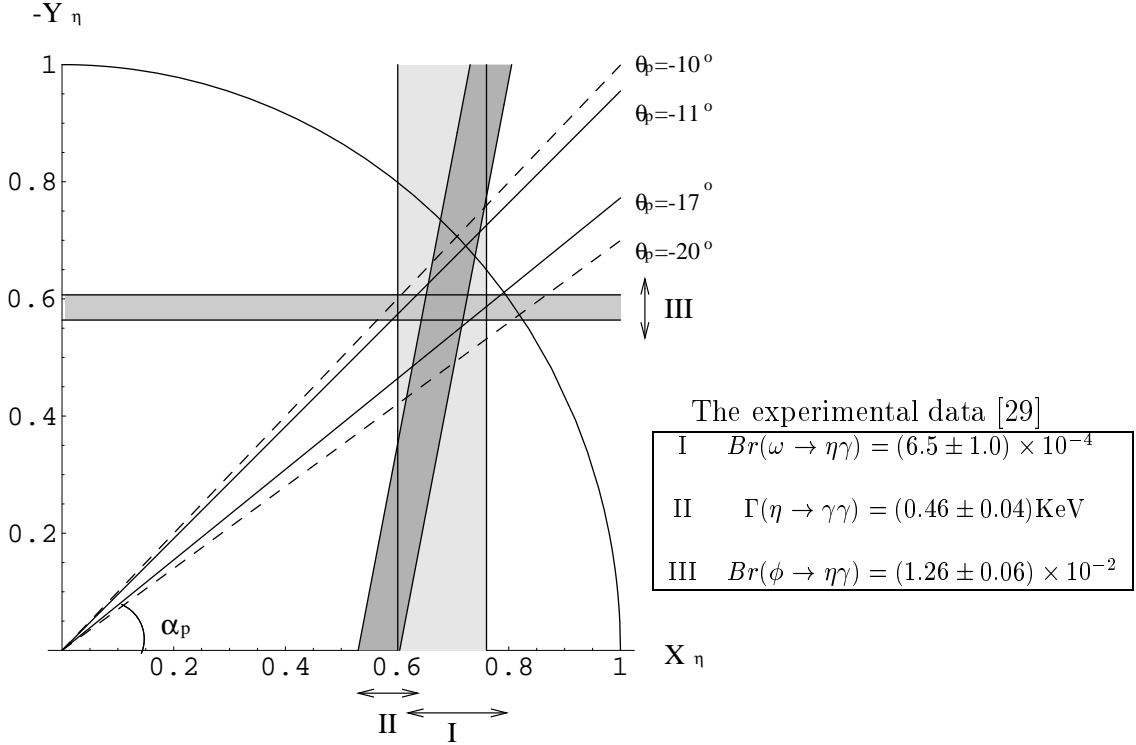}
 \end{center}
\vspace{-1cm}
 \caption[fig1]
{\vspace{-0.5cm} 
The experimental bounds for $\omega\to\eta\gamma$ (I), 
$\eta\to\gamma\gamma$ (II) and 
$\phi\to\eta\gamma$ 
\vspace{-0.5cm}
(III).  
The condition for $X_\eta$ and $Y_\eta$ in Eq. (\ref{X+Y=1}) 
is shown as a circumference. 
We 
obtain a constraint $-17^{\circ}<\theta_p<-11^{\circ}$. }
 \label{fig1}
\end{figure}

 First, we analyse $\omega\to\eta\gamma$, $\eta\to\gamma\gamma$ and 
$\phi\to\eta\gamma$ decays. 
Substituting the left hand side of 
Eq.(\ref{form1}) $\sim$ (\ref{form3}) 
for the experimental data and the errors ~\cite{pdg}, 
we obtain the constraint on $X_\eta$ and $Y_\eta$ and consequently, 
$\alpha_p$ via Eq. (\ref{defy}). 
The result is shown in Figure \ref{fig1}.  
The circumference denotes the constraint for $X_\eta$ and $Y_\eta$ 
in Eq. (\ref{X+Y=1}). 
As we estimated in the previous section, 
the theoretical error of 15\% is included.

\newpage 

In Figure \ref{fig1}, we have plotted simply the averages in  
the Review of Particle Physics ~\cite{pdg}. 
However, the experiments still have large errors for these processes. 
Looking carefully at the data in ~\cite{pdg}, we analyse the result 
depicted in Figure \ref{fig1}. 
A result for $\eta\to\gamma\gamma$ decay in 1974, 
$\Gamma(\eta\to\gamma\gamma)=(0.32\pm0.046)\mbox{KeV}$, is inconsistent with 
all other experiments so that we excluded this result when averaging. 
Consequently, the central value of $\Gamma(\eta\to\gamma\gamma)$ gets 
an increase of 5\%, which leads the bound II in Figure \ref{fig1} 
to shift to the right by about 0.03. 
After the shift, the bound II intersects the circle between 
$\alpha_p \simeq -44^{\circ}\mbox{and}-41^{\circ}$ 
and we obtain the result from 
the $\eta\to\gamma\gamma$ decay as 
$\theta_p\simeq -14^{\circ} \sim -11^{\circ}$. 
Similarly, a result for $\omega\to\eta\gamma$ in 1977, which is 
$Br(\omega\to\eta\gamma)=(3.0 {+2.5 \atop -1.8})\times10^{-4}$, 
is small compared to other data and in fact, it has a 70\% error. 
Exclusion of this value leads to a 6\% increase of the center value and 
about a 0.04 shift to the right of the bound I in Figure \ref{fig1}. 
As a result, 
the bound I intersects the circle at 
$\theta_p\simeq-17^{\circ}\sim-8^{\circ}$.
Finally, the experiment in 1983 of $\phi\to\eta\gamma$ reports a branching 
ratio $Br(\phi\to\eta\gamma)=(0.88\pm0.20)\times10^{-2}$ 
which is smaller than any other vlues. 
We exclude this result and obtain a 0.01 upward shift 
of the bound III in Figure \ref{fig1}. 
Then the result for $\theta_p$ from 
$\phi\to\eta\gamma$ is $-20^{\circ}\sim-11^{\circ}$. 

Eventually, we conclude that the experimental result for $\theta_p$ 
converges in a range of $-17^{\circ}\sim-11^{\circ}$. 
Note that we obtained a smaller value of $|\theta_p|$ than the previous work 
~\cite{etap1} which gave $-21^{\circ}<\theta_p<-16^{\circ}$. 
The change is mainly caused by two facts:  
the average of the decay rate of $\eta\to\gamma\gamma$ 
became smaller, and 
we utilized differently defined decay constants 
for $\eta$ and $\eta^{\prime}$. 


\subsection{Result for  $X_{\eta^{\prime}}$, $Y_{\eta^{\prime}}$ and 
$Z_{\eta^{\prime}}$  (determination of $Z_{\eta^{\prime}}$)}
\label{sec42}

\begin{figure}[tbh]
 \begin{center}
\vspace*{-4cm}
\includegraphics[\linewidth,\linewidth]{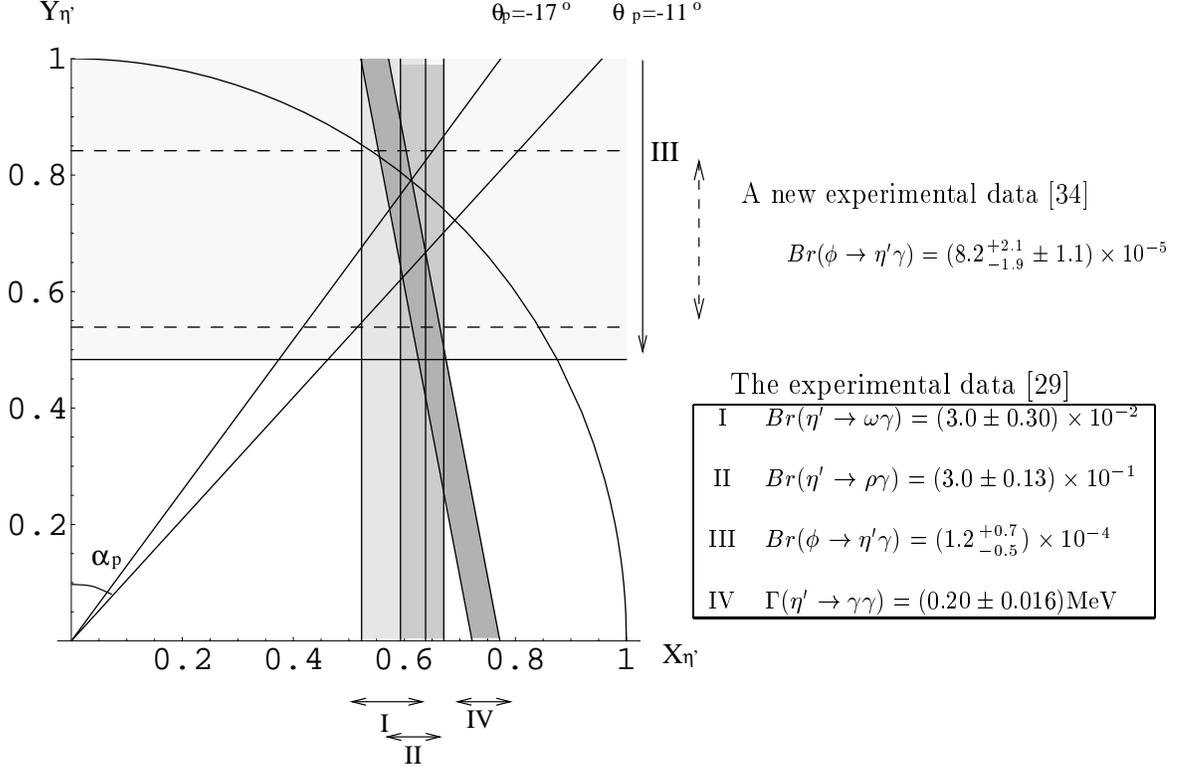}
 \end{center}
 \vspace{-1cm}
 \caption[fig2]
{\vspace{-0.5cm}
The experimental bounds for 
$\eta^{\prime}\to\omega\gamma$ (I), 
$\eta^{\prime}\to\rho\gamma$ (II), $\phi\to\eta^{\prime}\gamma$ 
\vspace{-0.5cm}(III)
and $\eta^{\prime}\to\gamma\gamma$ (IV). 
The dashed line shows the bound for $\phi\to\eta^{\prime}\gamma$ from 
a \vspace{-0.5cm} new experimental data 
[34]. 
Taking the value of $\theta_p$ to be $-11^{\circ}$ which is 
allowed \vspace{-0.5cm}in Figure \ref{fig1}, 
we observe the maximum 26\% of the gluonic component in $\eta^{\prime}$.}
 \label{fig2}
\end{figure}

Now we analyse 
$\eta^{\prime}\to\omega\gamma$, $\eta^{\prime}\to\rho\gamma$, 
$\eta^{\prime}\to\gamma\gamma$ and 
$\phi\to\eta^{\prime}\gamma$ decays. 
Constraints on 
 $X_{\eta^{\prime}}$, $Y_{\eta^{\prime}}$ and $Z_{\eta^{\prime}}$ 
can be obtained by 
using Eqs. (\ref{form4}) $\sim$ (\ref{form7}). 
The experimental bounds ~\cite{pdg} for 
these decays are shown in Figure \ref{fig2}.
As in the case of $\eta$, 
a 15 \% theoretical error is taken into account. 
From the analysis in Section \ref{sec41}, we have a constraint on 
$\theta_p$ between $-17^{\circ}$ and $-11^{\circ}$. 
Since we have a relation 
$X_{\eta^{\prime}}^2+Y_{\eta^{\prime}}^2+Z_{\eta^{\prime}}^2=1$, 
the result $X_{\eta^{\prime}}^2+Y_{\eta^{\prime}}^2<1$ represents  
$\eta^{\prime}$  having a gluonic component.

\noindent

\newpage

\vspace{0.5cm}
We have the following observations:  

\begin{quote}
$\star$
The maximum gluonic admixture in $\eta^{\prime}$ is obtained to be 
6\% for $\theta_p=-17^{\circ}$, 17\% for $\theta_p=-14^{\circ}$ 
and 26\% for $\theta_p=-11^{\circ}$ 
where the percentage is computed by 
\vspace*{-0.5cm}
\begin{equation}
R=\frac{Z_{\eta^{\prime}}}
{X_{\eta^{\prime}}+Y_{\eta^{\prime}}+Z_{\eta^{\prime}}}. 
\end{equation}
$\star$
If future experiments show an increase of 10\% in the central values of 
the $\eta^{\prime}\to\rho\gamma$ or 
$\eta^{\prime}\to\gamma\gamma$ decay rate, 
the existence of the gluonic content in $\eta^{\prime}$ 
will be excluded for large $|\theta_p|$. \\
$\star$
The CMD-2 collaboration observed $\phi \to \eta^{\prime} \gamma$ in 1999. 
Using their new result ~\cite{CMD-2} 
\vspace*{-0.5cm}
\[Br(\phi\to\eta^{\prime}\gamma)= 
(8.2{+2.1 \atop -1.9}\pm1.1)\times 10^{-5},\] 
the dashed bound in Figure \ref{fig2} is obtained. 
The new data show that the observation of the maximum gluonic admixture 
described above is still allowed. 
A more stringent constraint is expected once the data from 
the $\phi$ factory at DA$\Phi$NE come out. 
\end{quote}

\vspace*{-2cm}
\renewcommand{\baselinestretch}{2.0}

\vspace*{1cm}
\section{$J/\psi$ decays}

\begin{figure}[h]
 \begin{center}
\resizebox{10cm}{!}{\includegraphics{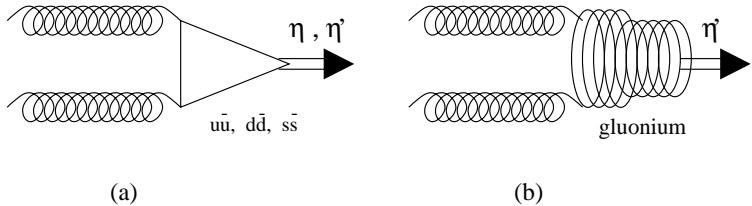}}
 \end{center}
\vspace*{-1cm}
 \caption[fig3]
{\vspace{-0.5cm}
Coupling of $\eta$ and $\eta^{\prime}$ to two gluons 
through quark and anti-quark triangle loop (a) and 
through gluonic admixture (b). }
 \label{fig3}
\end{figure}

Now we analyse the radiative $J/\psi$ decays into $\eta$ and $\eta^{\prime}$ 
and see the influence of the allowed amount of gluonic admixture 
in Section \ref{sec42} on the amplitudes. 
The ratio of the two decay rates $R_{J/\psi}$ can be written as 
~\cite{Kroll9,jpsi1,-,jpsi9}

\begin{equation}
  R_{J/\psi}
  =\frac{\Gamma(J/\psi\to\eta\gamma)}{\Gamma(J/\psi\to\eta^{\prime}\gamma)}
  = 
  \left(\frac{1-m_{\eta}^2/m_{J/\psi}^2}
  {1-m_{\eta^{\prime}}^2/m_{J/\psi}^2}\right)^3 
  |\frac{\sqrt{2} \xi X_{\eta}-\zeta(-Y_{\eta})}
  {(\sqrt{2} \xi X_{\eta^{\prime}}+ \zeta Y_{\eta^{\prime}})
  +g_r^{\prime} Z_{\eta^{\prime}}}|^2 \label{j2} 
\end{equation}
where $\xi$, $\zeta$ and $g_r^{\prime}$ are $f_{\pi}/f_x$,  
$f_{\pi}/f_y$, and the coupling of two gluons to gluonium, 
respectively.              
Using the average of ~\cite{pdg}, we have 
\begin{equation}
 R_{J/\psi}= \frac{\Gamma(J/ \psi \to \eta \gamma)}{\Gamma(J/ \psi \to
\eta^{\prime} \gamma)}
= 0.20 \pm 0.02.  \label{defr}
\end{equation}
The terms $\sqrt{2} \xi X_{\eta^{(\prime)}}$ and $\zeta Y_{\eta^{(\prime)}}$ 
in Eq.(\ref{j2}) represent the contributions from 
such intermediate processes as 
$gg \to$ ($u\bar{u}, d\bar{d}$ triangle loop) 
$\to \eta^{(\prime)}$ and 
$gg \to$ ($s\bar{s}$ triangle loop) $\to \eta^{(\prime)}$, respectively 
(see Figure \ref{fig3}(a)) and the term 
$g_r^{\prime}Z_{\eta^{\prime}}$ from 
$gg \to$ (gluonium) $\to \eta^{\prime}$ (see Figure \ref{fig3}(b)).
We define the ratio between the amplitudes for the process 
Figure \ref{fig3}(b) and Figure \ref{fig3}(a) by $r$: 
\begin{equation}
	r=\frac{g^{\prime}_rZ_{\eta^{\prime}}}
	{(\sqrt{2} \xi X_{\eta^{\prime}}+ \zeta Y_{\eta^{\prime}})}
\end{equation}

\vspace*{0.5cm}

\begin{figure}[thb]
 \begin{center}
\vspace*{-8.5cm}
\hspace*{2cm}
\includegraphics[\linewidth,\linewidth]{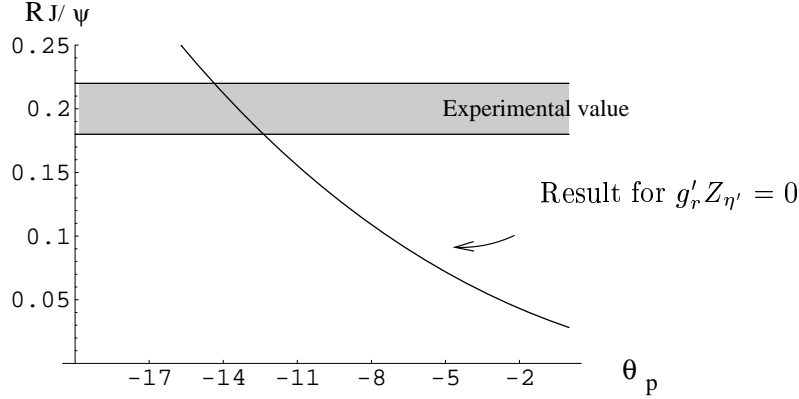}
 \end{center}
\vspace{-1cm}
 \caption[fig4]
{\vspace{-0.5cm}
The determination of $\theta_p$, 
putting $g_{r}^{\prime}Z_{\eta^{\prime}}=0$ 
(no gluonic admixture in $\eta^{\prime}$). 
The result conflicts with the 
observation in Section \ref{sec42} when $g_{r}^{\prime}\neq0$.  }
 \label{fig4}
\end{figure}

First, we examine the case of $r=0$ which means 
that gluonium does not contribute to 
$J/\psi\to\eta^{\prime}\gamma$ amplitude. 
In this case, 
the right hand side of Eq. (\ref{j2}) depends on only one parameter 
$\alpha_p$, so using Eq. (\ref{defr}), 
$\theta_p$ can be determined. 
The result is shown in Figure \ref{fig4}. 
We observe that for $g^{\prime}_rZ_{\eta^{\prime}}=0$, the $\theta_p$ 
angle is determined in a region $\theta_p=-13^{\circ}\pm1.0^{\circ}$. 
On the other hand, in the analysis of the glue content 
in Section \ref{sec42},  
$Z_{\eta^{\prime}}=0$ is allowed only when $\theta_p$ is in a 
narrow region around $-17^{\circ}$ (see Figure \ref{fig2}). 
This disagreement indicates that 
$Z_{\eta^{\prime}}=0$ should be excluded.

\vspace*{0.5cm}

\newpage

\begin{figure}[hb]
\vspace*{-8.5cm}
\hspace*{1.5cm}
 \begin{center}
\includegraphics[\linewidth,\linewidth]{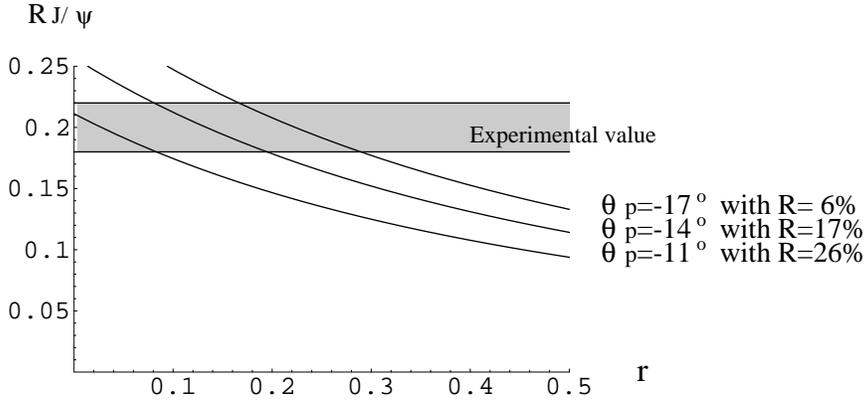}
 \end{center}
\vspace{-1cm}
 \caption[fig5]
{\vspace{-0.5cm} 
The amplitude 
of the process $J/\psi \to \eta^{\prime} \gamma $ has 
a maximum 20\% of contribution from gluonium in $\eta^{\prime}$ when 
we choose $\theta_p=-17^{\circ}$ with $R=6\%$. } 
 \label{fig5}
\end{figure}

Now let us examine the case of 
$g_r^{\prime}Z_{\eta^{\prime}}\neq0$ in Eq. (\ref{j2}). 
Since we do not know the value of $g_r^{\prime}$ which denotes 
the coupling of two gluons to gluonium we fix the $\theta_p$ angle at 
$-17^{\circ}$, $-14^{\circ}$ and 
$-11^{\circ}$ and examine each case. 
We set the value of $Z_{\eta^{\prime}}$ at the maximum which is 
allowed in Section \ref{sec42}. 
Substituting the left hand side of Eq. (\ref{j2}) for the experimental data, 
we determine the $r$ value for each $\theta_p$ angle. The result is shown in 
Figure \ref{fig5}. 
We observe that $r$ reaches a maximum of 0.3 
when $\theta_p \ \mbox{is} -17^{\circ}$ with 6\% of the glue content. That is, 
the amplitude of the process $J/\psi \to \eta^{\prime} \gamma $ has 
a maximum contribution of 20\% from gluonium in $\eta^{\prime}$.

%
%

\newpage
\renewcommand{\baselinestretch}{2.0}

\section{Conclusion}

We have examined the gluonic component of $\eta^{\prime}$ and 
the contributions to the process $gg \to \eta^{\prime}$. 
By analysing the latest experimental data  
on the radiative light meson decays, 
we have observed that 
the maximum 26 \% of the gluonic component in $\eta^{\prime}$ 
is possible at $\theta_p=-11^{\circ}$. 
Our constraint on the pseudoscalar mixing 
angle is $-17^{\circ}<\theta_p<-11^{\circ}$. 
Further investigation would be done once the data from 
DA$\Phi$NE will come out. 
We have also studied the contributions of gluonium  
to the radiative $J/\psi$ decays. 
Combining the obtained result from the analysis on the radiative light 
meson decays, we found that the  
$J/\psi$ decays also demand gluonium in $\eta^{\prime}$.  
In a case when we choose $\theta_p=-17^{\circ}$ with 6\% of gluonium 
in $\eta^{\prime}$, we have observed that the 20\% of the amplitude 
of $J/\psi\to\eta^{\prime}\gamma$ comes from gluonium. 

\newpage 
\vspace*{0.3cm}
\noindent
{\Large \textbf{Acknowledgments}}

The author would like to gratefully thank  Professor A. I. Sanda for 
suggesting that I examine all available experimental data which give 
information on the gluonic admixture of $\eta^{\prime}$,  
all his advises and encouragements through this work. 
The author is also grateful to Professor J. L. Rosner for 
fruitful discussions with respect to the new data of 
$\phi \to \eta^{\prime} \gamma$ and a critical reading of the manuscript. 
This work is supported by the Japanese Society for the 
Promotion of Science.


\vfill\eject

\end{document}